\documentclass[fleqn,twoside]{article}
\usepackage{espcrc2}

\usepackage{graphicx}
\usepackage[figuresright]{rotating}

\newcommand{\AmS}{{\protect\the\textfont2
  A\kern-.1667em\lower.5ex\hbox{M}\kern-.125emS}}
\newcommand{\bbox}[1]{\mbox{\boldmath$#1$}}

\title{A More Improved Lattice Action for Heavy Quarks}
\author{\underline{M. B. Oktay}\address[UIUC]{Department of Physics,
University of Illinois, 1110 West Green Street, Urbana, IL 61801}%
, 
        A. X. El-Khadra\addressmark[UIUC]\address[FNAL]{Theoretical
Physics Group, Fermi National Accelerator Laboratory, 
        P.O. Box 500, Batavia, IL 60510},
        A. S. Kronfeld\addressmark,
        P. B. Mackenzie\addressmark}

\begin{document}

\begin{abstract}
We extend the Fermilab formalism for heavy quarks to develop a more
improved action.
We give results of matching calculations of the improvement couplings
at tree level. Finally, we estimate the discretization errors associated
with the new action.
\end{abstract}

\maketitle

\section{INTRODUCTION}

One source of uncertainty in lattice QCD comes from the discretization
effects in numerical simulations.
Here we extend the Fermilab formalism to higher dimensional
operators to obtain a more improved action for heavy quarks, aiming to
reduce this uncertainty to 1--2\%.

The method starts with Symanzik~\cite{symanzik}, who introduced
a local effective Lagrangian
\begin{equation}
    \mathcal{L}_{\mathrm{eff}} = \mathcal{L}_{\mathrm{QCD}} +
	{\textstyle \sum_j} a^{\dim\mathcal{L}_j-4}K_j\mathcal{L}_j
    \label{eq:Sym}
\end{equation} 
of a continuum field theory.
The $\mathcal{L}_j$ are higher-dimensional operators.
The lattice action is adjusted so that the short-distance $K_j$
coefficients vanish, at least approximately.

The Fermilab formalism~\cite{elkhadra97} takes Wilson
fer\-mions~\cite{wilson} with the clover action~\cite{clover}
and introduces different couplings for space-like and time-like
interactions.
Ref.~\cite{elkhadra97} considers operators up to dimension five. 
Ref.~\cite{oktay02} adds dimension-six operators to this action.
In this paper, we give a more complete action which has dimension
six and some dimension seven operators and determine the couplings by
matching at tree level.

\section{QUARK BILINEARS} 

In this section, we consider the quark bilinears.
The dimension-five (-six) interactions are listed in
Ref.~\cite{elkhadra97} (Ref.~\cite{oktay02}).
Redundant directions are identified by field transformations.
They are used to eliminate the higher-order time derivatives,
which would give large errors for large masses.
We do not repeat this analysis here.

The corrections in Eq.~(\ref{eq:Sym}) should be classified by a
power-counting scheme.
For light quark physics the only relevant ratio is $\Lambda a$.
For heavy quark physics, however, the power counting should follow
HQET for heavy-light systems or NRQCD for quarkonium.
In HQET, operators are assigned a power of $\lambda=\Lambda/m_Q$;
in NRQCD, a power of relative velocity~$v$.
In these power-counting schemes, some dimension-seven interactions
are commensurate with dimension-six interactions.
As seen in the list in Table~\ref{table:1},
they scale as $\lambda^3$ or~$v^6$.
\begin{table}[tb]
\caption{Dimension six and seven bilinear interactions in
$\mathcal{L}_{\rm eff}$ with HQET and NRQCD power counting.}
\label{table:1}
\newcommand{\m}{\hphantom{$-$}}
\newcommand{\cc}[1]{\multicolumn{1}{c}{#1}}
\renewcommand{\tabcolsep}{0.35pc} 
\renewcommand{\arraystretch}{1.3} 
\begin{tabular}{@{}lccc}
\hline
dim  & Operator  &  HQET & NRQCD \\
\hline
6 & $\bar{Q}\gamma_iD_i^3\bar{Q}$ & $\lambda^3$ & $v^4$ \\
  & $\bar{Q}\bbox{\gamma}\cdot\bbox{D}Q$ & $\lambda^3$ & $v^4$ \\
  & $\bar{Q}\{\bbox{\gamma}\cdot\bbox{D},\bbox{\alpha}\cdot\bbox{E}\}Q$
    & $\lambda^2$ & $v^4$ \\
  & $\bar{Q}\{\bbox{\gamma}\cdot\bbox{D}, i\bbox{\Sigma}\cdot\bbox{B}\}Q$
    & $\lambda^3$ & $v^6$ \\
  & $\bar{Q}[D_4,\bbox{\gamma}\cdot\bbox{E}]Q$ & $\lambda^4$ & $v^6$ \\
  & $\bar{Q}\gamma_4(\bbox{D}\cdot\bbox{E}-\bbox{E}\cdot\bbox{D})Q$
    & $\lambda^2$ & $v^4$ \\
\multicolumn{2}{r}{$\bar{Q}\bbox{\gamma}\!\cdot\!(\bbox{D}\!\times\!\bbox{B}+
\bbox{B}\!\times\!\bbox{D})Q$} & $\lambda^3$ & $v^6$ \\
\hline
7 & $\bar{Q}(\bbox{D}^2)^2Q$ & $\lambda^3$ & $v^4$ \\
& $\bar{Q}D_i^4Q$         & $\lambda^3$ & $v^4$ \\
& $\bar{Q}\{\bbox{D}^2, i\bbox{\Sigma}\cdot\bbox{B}\}Q$ & $\lambda^3$ & $v^6$ \\
& $\bar{Q}\bbox{\gamma}\cdot\bbox{D}i\bbox{\Sigma}\cdot\bbox{B}
\bbox{\gamma}\cdot\bbox{D}Q$
& $\lambda^3$ & $v^6$ \\
& $\sum_{i\neq j}\bar{Q}\{D_j^2,i\Sigma_iB_i\}Q$ & $\lambda^3$ & $v^6$ \\
& $\sum_{i\neq j}\bar{Q}i\Sigma_iD_jB_iD_jQ$ & $\lambda^3$ & $v^6$ \\
& $\bar{Q}D_ii\bbox{\Sigma}\cdot\bbox{B}D_iQ$ & $\lambda^3$ & $v^6$ \\ 
& $\bar{Q}\bbox{D}\cdot(\bbox{B}\times\bbox{D})Q$ & $\lambda^3$ & $v^6$ \\
& $\bar{Q}\bbox{E}\cdot\bbox{E}Q$ & $\lambda^3$ & $v^6$ \\
& $\bar{Q}i\bbox{\Sigma}\cdot(\bbox{E}\times\bbox{E})Q$ & 
$\lambda^3$ & $v^6$ \\
& $\bar{Q}\bbox{B}\cdot\bbox{B}Q$ & $\lambda^3$ & $v^8$ \\
& $\bar{Q}i{\Sigma}\cdot(\bbox{B}\times\bbox{B})Q$ &  
$\lambda^3$ & $v^8$ \\
\hline
\end{tabular}
\end{table}
Further dimension-seven operators carry power $\lambda^4$ (or higher)
in HQET and $v^8$ (or higher) in NRQCD.

Including the needed dimension-seven operators,
our lattice action takes the form
\vspace*{-0.8pt}
\begin{eqnarray}
    S_F & = & S_0+S_B+S_E
\nonumber \\
    & + & a^2c_1 \int\bar\psi\{\bbox{\gamma}\!\cdot\!\bbox{D},
        \bbox{D}^2\}\psi
    + a^2c_2 \int \bar\psi\gamma_iD_i^3\psi
\nonumber \\
    & + & a^3r_3\int\bar\psi(\bbox{D}^2)^2\psi
    +a^3r_4\int\bar\psi D_i^4\psi
\nonumber \\
    & + & a^2c_5\int\bar\psi\{\bbox{\gamma}\cdot\bbox{D},
	i\bbox{\Sigma}\cdot\bbox{B}\}\psi
\nonumber \\
     & + & a^2c_6\int\bar\psi(\bbox{D}\cdot
    \bbox{E}-\bbox{E}\cdot\bbox{D})\psi
\nonumber \\
    & + & a^2c_7\int\bar\psi
        \bbox{\gamma}\cdot(\bbox{D}\times\bbox{B}+
    \bbox{B}\times\bbox{D})\psi
\nonumber \\
    & + & a^2c_8\int\bar\psi\{\bbox{\gamma}\cdot\bbox{D},
        \bbox{\alpha}\cdot\bbox{E}\}\psi
\nonumber  \\
    & + & a^3r_9\int\bar\psi\{\bbox{D}^2,i\bbox{\Sigma}\cdot\bbox{B}\}\psi
\nonumber \\
    & + & a^3r_{10}\int\bar\psi
        {\textstyle\sum_{i\neq j}}\{i\Sigma_iB_i,D_j^2\}\psi
\nonumber \\
    & + & a^3r_{11}\int\bar\psi
        {\textstyle\sum_{i\neq j}}i\Sigma_iD_jB_iD_j\psi
\nonumber \\
    & + & a^3r_{12}\int\bar\psi
        \bbox{\gamma}\cdot\bbox{D}i
	\bbox{\Sigma}\cdot\bbox{B}\bbox{\gamma}\cdot\bbox{D}\psi ,
\label{eq:S}
\end{eqnarray}
where conventional difference operators and clover fields are
understood.
We have included some terms that we expect to be redundant, to retain
some flexibility to, say, speed up computer simulations.

For brevity we do not introduce terms, corresponding to the last six
rows of Table~\ref{table:1}.
The operators $\bar{Q}D_ii\bbox{\Sigma}\cdot\bbox{B}D_iQ$ and
$\bar{Q}\bbox{D}\cdot(\bbox{B}\times\bbox{D})Q$ are redundant.
Matching of the last four rows in Table~\ref{table:1} requires a
calculation of Compton scattering, which is is still in progress.

\section{MATCHING}

We derive matching conditions for the improvement couplings in
Eq.~(\ref{eq:S}) by matching on-shell quantities for small momentum
but any~$m_Qa$.
We do this by calculating the energy momentum relation and
the scattering of a heavy quark off a background field.
The energy momentum relation has been studied \cite{elkhadra97}
and summarized \cite{oktay02} before.
It gives us conditions on~$c_1$, $c_2$, $r_3$ and~$r_4$.
Results for the chromoelectric terms $c_E$, $c_6$ and $c_8$ are given
in Ref.~\cite{oktay02}, determined from the temporal component of the
background field.
Scattering from the spatial component of the background field yields
results for the chromomagnetic terms.
Our results yield the following matching conditions.
The energy momentum relation gives: 
\begin{eqnarray}
\lefteqn{
16\zeta c_1={4\zeta^4(1-\zeta^2)\over [m_0(2+m_0)]^2}-{3r_s^2\zeta^4\over 
(1+m_0)^2}
}  \nonumber \\
&+&{\zeta^3[2\zeta+4r_s(1+m_0)-6r_s\zeta^2/(1+m_0)]\over m_0(2+m_0)} 
\nonumber \\
&+&{m_0(2\!+\!m_0)\over 2(1\!+\!m_0)}
[{r_s^2\zeta^2\over (1\!+\!m_0)}\!-\!{r_s^3\zeta^3
\over (1\!+\!m_0)^2}\!-\!{8r_3}].
\end{eqnarray}
where $r_3$ can be chosen to be zero.
Also
\begin{equation}
    0 = \zeta^2+6\zeta c_2+(r_s\zeta-24r_4){m_0(2+m_0)\over 8(1+m_0)}.
\end{equation}
We do not set $r_4=0$ or $r_s\zeta/24$ but choose it below so that
$r_{11}=0$.
For the chromoelectric interactions one finds
\begin{eqnarray}
\zeta^2c_E^2+c_8{2m_0(2\!+\!m_0)\over (1\!+\!m_0)} & = &
{\zeta^2(\zeta^2\!-\!1)\over m_0(2\!+\!m_0)} 
\!+\!{r_s\zeta\over (1\!+\!m_0)} \nonumber \\
&+&{r_s^2m_0(2+m_0)\over 4(1+m_0)^2},
\end{eqnarray}
\begin{equation}
c_6=0.
\end{equation}
For the chromomagnetic interactions one finds
\begin{eqnarray}
c_B&=&r_s, \quad
c_7 = 0, \quad
c_5=c_1, \quad
r_9 = r_3, \\
r_{10}&=&-{1\over 48}(r_s\zeta-24r_4)+{1\over 12}c_B\zeta, \\
r_{11}&=&-{1\over 48}(r_s\zeta-24r_4)-{1\over 6}c_B\zeta, \\
r_{12}&=&0.
\end{eqnarray}
In summary, we may set $c_6=c_7=c_8=r_3=r_9=r_{11}=r_{12}=0$, leaving only five
non-zero couplings: $c_1$, $c_2$, $r_4$, $c_5$ and~$r_{10}$.

\section{ERROR ESTIMATES}
We now estimate the discretization effects.
It is advantageous to use a heavy-quark description of
cutoff effects~\cite{askHQET}, because then contributions from, say,
$\bar{Q}\bbox{\alpha}\cdot\bbox{E}Q$ and
$\bar{Q}\{\bbox{\gamma}\cdot\bbox{D},\bbox{\alpha}\cdot\bbox{E}\}Q$
are automatically combined.
Thus, we express the uncertainties as
\begin{equation}
    \mathtt{Error}_i = \left(C_i^{\mbox{lat}}-C_i^{\mbox{cont}}\right)
        \langle O_i \rangle
\end{equation}
where $C_i$ are HQET/NRQCD short-distance coefficients for lattice gauge
theory and continuum QCD~\cite{askHQET}.
The mismatched coefficients can be obtained from the results of
the previous section.

\begin{figure}[bt]
\includegraphics*[height=5.75cm,width=7.5cm]{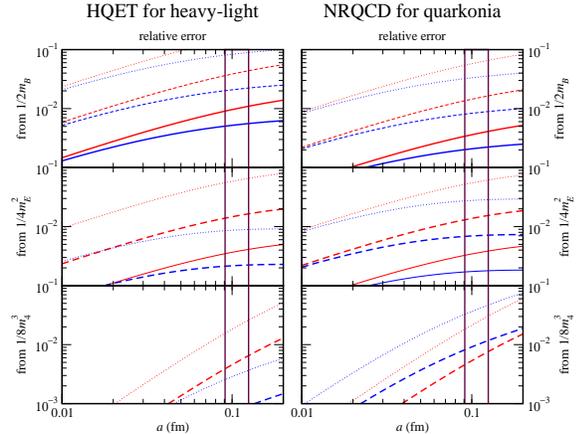}
\vspace*{-12pt}
\caption{Truncation errors for the new lattice action. 
The dotted curves correspond to the unimproved case.
The dashed and solid curves show the errors from tree-level
and one-loop matching, respectively.}
\label{fig:errors}
\vspace*{-18pt}
\end{figure}

The matrix elements $\langle O_i \rangle$ are estimated by using HQET
(NRQCD) power counting for heavy-light systems (quarkonia).
Since higher-order calculations are not yet available we multiply the
tree-level mismatch with $\alpha_s^l$ to estimate the $l-$loop mismatch.
This is a conservative estimate because the most pessimistic asymptotic
behavior for the coefficients of the improvement operators is the same
at higher orders as in the tree level formulas of previous section.
To obtain the numbers shown in Figure~\ref{fig:errors}, we
take $\alpha_s=0.25$.

Fig.~\ref{fig:errors} shows results for
the chromomagnetic energy (labelled $1/2m_B$),
chromoelectric energy (labelled $1/4m_E^2$),
and the relativistic correction term (labelled $1/8m_4^3$).
See Refs.~\cite{elkhadra97,askHQET} for details.
The vertical lines represent $a=1/8$~fm and $a=1/11$~fm,
corresponding to the unquenched MILC lattices~\cite{MILC}.
The black (gray) curves represent the charmed (bottom) quark. 
Note that, in most cases, bottom quarks have smaller discretization
effects than charmed quarks.
Thus, Fig.~\ref{fig:errors} shows concretely that the Fermilab formalism
avoids errors of order~$(m_Qa)^n$.

\section{CONCLUSIONS}
We have presented a more improved action for heavy quarks with five
new operators.
Our error analysis shows that, to reduce errors below 1\%, one-loop
matching is essential for the coupling~$c_B$ of the chromomagnetic
clover term.
Tree-level matching for the coupling of the chromoelectric term, $c_E$,
is usually enough, but one-loop may be necessary for charmonium.
Tree-level matching suffices for the other couplings in
Eq.~(\ref{eq:S}).

M.B.O. and A.X.K. are supported in part by the U.S. Department of Energy
under contract DE-FG02-91ER40677.
Fermilab is operated by the Universities Research Association Inc.,
under contract with the U.S. Department of Energy.

\end{document}